\documentclass{ws-procs9x6}

\def\etal {{\it et al.}}

\newcommand{\om}{\omega}

\newcommand{\be}{\begin{equation}}
\newcommand{\ee}{\end{equation}}
\newcommand{\bea}{\begin{eqnarray}}
\newcommand{\eea}{\end{eqnarray}}
\begin{document}

\title{MATTER-WAVE TESTS OF THE\\
 GRAVITATIONAL REDSHIFT IN SPACE}

\author{H.\ M\"ULLER$^*$ and M.A.\ HOHENSEE}

\address{Department of Physics, University of California\\
366 Le Conte Hall, Berkeley, CA 94720, USA\\
$^*$E-mail: hm@berkeley.edu}

\author{N.\ YU}

\address{Jet Propulsion Laboratory,
California Institute of Technology\\
4800 Oak Grove Drive, Pasadena, CA 91109, USA}

\begin{abstract}
A recent measurement of the gravitational redshift was based on interference of matter waves.
Operation in microgravity can improve it by a factor of $10^5$ 
and, in some models, even $10^{10}$.
\end{abstract}

\bodymatter

\section{Introduction}

One of the key consequences of the Einstein Equivalence Principle is the gravitational redshift. Precision measurements of this effect are cornerstones of our trust in the theory of general relativity. For example, Pound, Rebka and Snyder have used stationary M\"ossbauer sources in a tower\cite{PoundRebkaSnyder} and verified the redshift within an accuracy of about 1\%. The Gravity-Probe A (GP-A) experiment consisted of a hydrogen maser launched to a height of 10,000\,km in a sounding rocket\cite{Vessot} and obtained an accuracy of $7\times 10^{-5}$.

\section{Atomic physics tests}

Recently, a 10,000 fold improvement has been reached by atom interferometry\cite{redshift}, see Fig. 1: an anomalous gravitational redshift would produce a measurable phase shift $\phi$ between the interferometer arms by modifying the Compton frequency of matter waves. Here, we present this experiment and propose a space-based version of it.

The action of a point particle in general relativity is given by $
S=\int mc^2d\tau$, where $d\tau=\sqrt{-g_{\mu\nu}dx^\mu dx^\nu}$ is the proper time interval. These expressions include the gravitational redshift and the special relativistic time dilation for a moving clock. In the semiclassical limit, we obtain\cite{Borde5D}
\be\label{Comptonphase}
\phi=\frac i\hbar \int mc^2d\tau=\int \om_Cd\tau,
\ee
where $\om_C$ is the Compton frequency. From this, it is clear that many effects in quantum mechanics are connected to the gravitational redshift and special relativistic time dilation.

\subsection{`Pound-Rebka like' tests: cold atoms in optical lattices}

Bloch oscillations of cold atoms in accelerated optical lattices\cite{Peik} are a well known effect in atomic physics. In the experiments of interest here, a vertical standing wave of light causes a periodic potential with potential minima spaced by half the laser wavelength, $\lambda/2$.

According to Eq. (\ref{Comptonphase}), the atomic states located in neighboring potential minima have their Compton frequencies redshifted relative to each other. When the atoms are released from the optical lattice (by abruptly switching off the standing wave, for example), the states located in different lattice sites interfere. This interference is governed by a beat frequency
\be
\om_B=(1+\beta)\frac\lambda 2 \frac{g}{c^2}\frac{mc^2}{\hbar},
\ee
which is commonly known as the Bloch frequency. We have included a factor of $1+\beta$ to account for a possible redshift anomaly. It can be observed as an oscillation of the atom's velocity over time.

A measurement with rubidium atoms\cite{Clade} leads to $\beta =(3\pm1)\times 10^{-6}$. Guglielmo Tino (private communication) has improved the accuracy to a level of approximately $|\beta|\lesssim 10^{-7}.$ We conclude that Pound-Rebka like matter wave redshift measurements lead to limits of a few parts in $10^6$ at present. This is already an improvement by a factor of more than 10 compared to the best classical tests.l

\subsection{`GP-A like' tests: atom interferometers}

The Mach-Zehnder atom interferometer (Fig. 1) provides us with an analogy to GP-A in atomic physics: like in the original GP-A, we obtain a larger separation of the clocks and hence a higher sensitivity. At the same time, however, the atoms are moving. Thus, corrections for the special relativistic time dilation will have to be applied in order to extract the gravitational redshift.

\begin{figure}
\centering
\psfig{file=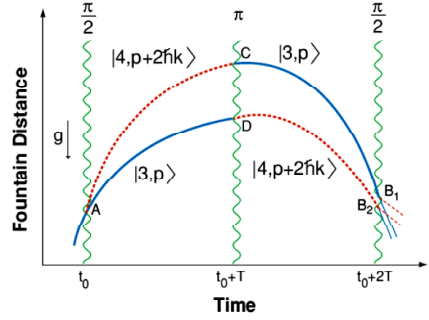,width=0.55\textwidth}
\caption{Mach-Zehnder atom interferometer. A laser-cooled atom in free fall is subjected to three pulses from a pair of anti-parallel laser beams having respective wavenumbers of $k_1$ and $k_2$.
The first laser pulse transfers the momentum $\hbar (k_1+k_2)$ two photons to the atom, with a probability of 50\%. As a result, it places the atom into a coherent superposition of two quantum states, which physically separate owing to
their relative momentum $k$. The second pulse redirects the atom momentum so that the paths merge and interfere at the time of the third pulse. }
\end{figure}

The atom interferometer contains a built-in mechanism for monitoring the trajectories and applying this correction: whenever a photon is absorbed by the atom, the photon's phase is added to the phase of the matter wave. For stimulated emission of a photon, the opposite occurs. The photon's phase $kz+\phi_0$, in turn, is a function of where the interaction occurs (where $\phi_0$ is its value at some reference location). Summing up the photon's phases over all interactions (Fig. 1), we obtain $\phi_{\rm i}=+kg'T^2$.
This contains the actual acceleration of free fall $g'$ (not necessarily equal to $g$ that enters Eq.\ (\ref{Comptonphase})) because it is a function of where the atom is when it interacts with the photon. The final result for the atom interferometer phase can be written as
\be\label{MZphase}
\Delta\phi=\underbrace{(1+\beta)kgT^2-kg'T^2}_{\Delta \phi_{\rm free}}+kg'T^2.
\ee
where $\Delta \phi_{\rm free}$ is the result of integrating Eq. (\ref{Comptonphase}) and where we included a factor of $1+\beta$ to account for possible redshift anomalies. Thus, it turns out that the atom-light interaction phase cancels the one due to special relativistic time dilation. The acceleration of free fall $g'$ does not need to be known {\em a priori}. This is the analogy of measuring the trajectory and subtracting time dilation in classical redshift tests.

The most accurate data to date comes from an interferometer using caesium atoms
in an atomic fountain. An absolute gravimeter (an FG-5 falling corner cube gravimeter) was used close by and corrected for systematic effects, such as elevation, air pressure, tides and polar motion. The result of $\beta=(7\pm7)\times 10^{-9}$ was obtained.

\section{Measuring the redshift in microgravity}

The above atom interferometry measurement of the gravitational redshift was limited by the accuracy to which local gravity $g$ could be measured. Thus, space-based operation can improve the experiment by several orders of magnitude, as the ordinary effects of gravity vanish in a freely falling platform, whereas a redshift anomaly will still produce a signal. This removes the major limiting influence, the knowledge of local gravity. A redshift violation of $\beta\sim 10^{-14}$ can produce several microradian of phase shift between the interferometer arms, which can be measured at the shot noise level within one week of integration time.

For the removal of systematic effects, it is important to note that the frame of reference of the atom interferometer can be defined by using a retroreflection mirror to make the counterpropagating laser beams. It is the motion of this mirror that defines the frame of reference to which the atom interferometer signal is referred. Vibrations and residual accelerations of the rest of the setup are then unimportant.

Table 1 shows some gravitational and systematic effects in a spaceborne redshift experiment. The largest uncertainties are residual acceleration and the gravity gradient. These can be nulled using accelerometers and gradiometers developed, e.g., for the LISA, GOCE, and STEP\cite{Step} missions. The retroreflection mirror could, for example, be floating inside the craft, and the platform's trajectory could be servoed to the mirror. Magnetic fields will have to be suppressed to the microgauss level, which can be done by shielding, or by a double-interferometer wherein half the atoms operate as a magnetometer. The systematic effects are compatible with reaching a sensitivity of $10^{-14}$ in the redshift parameter, a 100,000 fold improvement relative to the best Earth-based experiment.

\begin{table}
\tbl{Gravitational effects in a space-borne atom interferometer. We assume a redshift anomaly of $\beta=10^{-14}$, Cs atoms in the $m_F=0$ state, a wavelength of 852\,nm, an orbit of 340\,km, $n=10$-photon beam splitters, same internal states, a pulse separation time $T=10\,$s, an initial atom velocity $v_0=1\,\mu$m/s, and a residual gravitational acceleration $g'=10^{-14}\,$m/s$^2$. `DK$n$' means line $n$ in Table 1 of Ref.\ \refcite{Dimopoulos}.}
{\begin{tabular}{@{}cc@{}}\toprule
Effect & Phase (rad) \\\colrule
Redshift anomaly & $6.5\times 10^{-5} (\beta/10^{-14}) (n/10) (T/10{\rm ~s})^2$ \\ Residual acceleration & $7.3\times 10^{-5} (n/10) (T/10 {\rm ~s})^2 (g'/10^{-14}{\rm ~m s}^{-2})$ \\ Gravity gradient & $-1.8\times 10^{-1} (n/10) (T/10{\rm ~s})^3 (v_0/\mu{\rm ~m s}^{-1})$ \\ Gravity gradient & $-1.1\times 10^2 (n/10) (T/10{\rm ~s})^4 (g'/10^{-5}{\rm ~m s}^{-2})$ \\ Finite speed of light & $-7.3\times 10^{-8} (n/10) (T/10{\rm ~s})^3$ \\ Doppler effect & $-7.4\times 10^{-10} (n/10) (T/10 {\rm ~s})^2 (v_0/\mu{\rm ~m s}^{-1}) (g'/10^{-5}{\rm ~m s}^{-2})$ \\ First gradient recoil & $-3.2\times 10^3 (n/10)^2 (T/10 {\rm ~s})^3$ \\ DK9 & $-1.3\times 10^{-5} (n/10)^2 (T/10{\rm ~s})^2 (g'/10^{-5}{\rm ~m s}^{-2})$ \\ Raman splitting DK13$^a$ & $-1.3\times 10^{-13} (n/10) (T/10{\rm ~s})^2 (f_{\rm HFS}/9 {\rm ~GHz}) (g'/10^{-5}{\rm ~m s}^{-2})$ \\ $g^3$ DK17 & $-5.7\times 10^{-20} (n/10) (T/10{\rm ~s})^4 (g'/10^{-5}{\rm ~m s}^{-2})$\\ Gravity gradient DK25 & $-7.2\times 10^{-9} (n/10)^2 (T/10{\rm ~s})^4 (g'/10^{-5}{\rm ~m s}^{-2})$ \\ Raman splitting DK26$^a$ & $2.3\times 10^{-11} (n/10) (T/10{\rm ~s})^2 (f_{\rm HFS}/9 {\rm ~GHz}) (g'/10^{-5}{\rm ~m s}^{-2})$  \\Shot noise & $10^{-6}({\rm ~flux}/10^6{\rm ~s}^{-1})^{-1/2}({\rm time}/{\rm week})^{-1/2}$ \\
Magnetic fields & $1.6\times 10^{-2}(n/10)(T/10{\rm ~s})^2(BdB/dz/({\rm mG})^2{\rm ~m}^{-1})$\\
Cold collisions$^a$ & $10^{-4}({\rm ~density}/10^6{\rm ~cm}^{-3})({\rm balance}/0.1) $\\\botrule
\end{tabular}}
\vskip 3 pt
$^{\text a}$ zero for same internal states.\\
\label{aba:tbl1}
\end{table}

According to a toy model for redshift violations\cite{MikeProc}, measurable redshift violations scale with the squared velocity $v^2/c^2$ of the clocks (here: matter-wave oscillations) relative to the source of the gravitational field. This is equal only a few m/s in ground-based experiments, but several km/s in a space experiment. For an experiment in a sounding rocket or a satellite in a highly elliptical orbit, an additional factor of improvement of $\sim 10^6$ is thus provided by the motion of the space platform.

\section*{Acknowledgments}

We thank S. Chu, S. Herrmann, J. Kollmeier, S.-Y. Lan, S. Ospelkaus, A. Peters, J. Phillips, and G. Tino for discussions. Support from the David and Lucile Packard Foundation, the Alfred P. Sloan Foundation and the National Institute of Standards and Technology is gratefully acknowledged.

\end{document}